\documentclass[preprint,showkeys,preprintnumbers,amsmath,amssymb]{revtex4}

\usepackage{CJK}
\usepackage{graphicx}
\usepackage{dcolumn}
\usepackage{bm}
\usepackage{hyperref}
\usepackage{multirow}
\usepackage{rotating}
\usepackage{url}
\usepackage{color}

\usepackage{amsfonts,amssymb,amsmath}
\usepackage{natbib}
\usepackage{bbm}
\usepackage{chemformula}
\usepackage{lineno}
\usepackage{booktabs}
\usepackage{float}
\usepackage{tcolorbox}
\usepackage[normalem]{ulem}

\newcommand{\delete}[1]{\bgroup\markoverwith{\textcolor{red}{\rule[0.5ex]{2pt}{1pt}}}\ULon{#1}}

\allowdisplaybreaks

\begin{document}

\title{Influence of particle geometry on dispersion force}

\author{Yifei Liu\footnote{yfliu@my.swjtu.edu.cn}}
\affiliation{Shenzhen Key Laboratory of Deep Underground Engineering Sciences and Green Energy, Shenzhen University, Shenzhen 518060, China}
\affiliation{Guangdong Provincial Key Laboratory of Deep Earth Sciences and Geothermal Energy Exploitation and Utilization, College of Civil and Transportation Engineering, Shenzhen University, Shenzhen 518060, China}
\author{Heping Xie}
\affiliation{Shenzhen Key Laboratory of Deep Underground Engineering Sciences and Green Energy, Shenzhen University, Shenzhen 518060, China}
\affiliation{Guangdong Provincial Key Laboratory of Deep Earth Sciences and Geothermal Energy Exploitation and Utilization, College of Civil and Transportation Engineering, Shenzhen University, Shenzhen 518060, China}
\author{Cunbao Li\footnote{cunbao.li@szu.edu.cn}}
\affiliation{Shenzhen Key Laboratory of Deep Underground Engineering Sciences and Green Energy, Shenzhen University, Shenzhen 518060, China}
\affiliation{Guangdong Provincial Key Laboratory of Deep Earth Sciences and Geothermal Energy Exploitation and Utilization, College of Civil and Transportation Engineering, Shenzhen University, Shenzhen 518060, China}
\author{Dong-Sheng Jeng}
\affiliation
{School of Engineering and Built Environment, Griffith University Gold Coast Campus, Queensland 4222, Australia}
\author{Bo Nan Zhang}
\affiliation{School of Nuclear Science and Technology, Lanzhou University, Lanzhou 730000, China}

\date{\today}

\begin{abstract}
Dispersion forces (van der Waals force and Casimir force) originating from quantum fluctuations are crucial in the cohesion of microscale and nanoscale particles.
In reality, these particles have a variety of irregular shapes that differ considerably from any idealized geometry.
Previous experiments have demonstrated that dispersion forces strongly depend on the geometry.
Because of the nonadditivity of these forces, commonly used numerical additive methods, such as the Hamaker and Derjaguin approximations, are not suitable for calculations with complex geometries.
Moreover, experimental studies are difficult to identify the contributions of the dispersion force from the many forces that constitute the cohesion.
Therefore, no general law about the influence of particle geometry on dispersion forces has been established.
Thus, in this paper, the fluctuating surface current (FSC) technique, an exact scattering theory-based nonadditive algorithm, was used to study this influence.
To characterize complex geometries, a data-adaptive spatial filtering method was introduced to perform scale decomposition, and descriptors at three observation levels (global, local, and surface) were used.
Based on the advanced geometric analyses and accurate numerical calculations, the influence of multiscale surface fluctuations on dispersion forces was determined. Furthermore, a convenient formula for predicting the dispersion forces between particles with complex shapes from the exact Lifshitz solution was established via multistage corrections.
\end{abstract}



\maketitle

\section{Introduction}
\label{sec:introduction}
Cohesion plays a key role in determining the behaviors of particles and particle systems ranging from the nanoscale to the microscale. The dispersion force is one of the main sources of cohesion. In nearly dry and uncharged particle systems with particle sizes less than 10 micrometers, dispersion forces dominate \citep{Israelachvili:1977intermolecular, Rimai:2000adhesion, Castellanos:2005relationship, Li:2006london}.

In different historical stages, the dispersion force has had various names. The earliest phenomenological name of the dispersion force was the 'van der Waals force' \citep{Van:1873over}, followed by the 'London force' \citep{London:1930theorie} and 'Casimir force' \citep{Casimir:1948influence}. However, in the 1950s, Lifshitz realized that these forces all originated from quantum fluctuations, and a unified theory was developed \citep{Lifshitz:2013statistical}. This theory can be applied to reproduce the van der Waals force (vdW force) and Casimir force as limiting cases of small and large separations \citep{Klimchitskaya:2015casimir}.
In this study, we use the general term `dispersion force' to refer to these forces with the same origin \citep{Buhmann:2013dispersion,Svetovoy:2015influence}.
In reality, almost all particles have various irregular shapes that differ considerably from any idealized geometry, including both natural and artificial particles.
Moreover, it has been proven both theoretically and experimentally that dispersion forces depend strongly on geometry, and changes in dispersion forces caused by geometric changes can reach several orders of magnitude. \citep{Montgomery:2000analytical, Emig:2001probing, Rodriguez:2011casimir, Wang:2021strong}.
Therefore, it is of significance to quantify the influence of particle geometries on dispersion forces.

For objects with arbitrary shapes, including objects made from idealized materials, the dispersion force is challenging to calculate analytically \citep{Chernodub:2020casimir}.
At present, the most effective methods for determining the influence of geometry on the dispersion force are physical experiments and numerical calculations.
Different experimental approaches have been developed in fields where these forces play important roles.
Many experimental studies are based on two main types of measuring equipment: the surface force apparatus (SFA) and atomic force microscope (AFM).
The SFA technique was developed by \citet{Israelachvili:2011intermolecular} and can be applied to measure the force between two macroscopically curved surfaces over relatively large areas with angstrom resolution.
Some researchers have studied the effect of surface roughness on contact mechanics with SFA approaches \citep{Benz:2006deformation,Valtiner:2011effect, Dziadkowiec:2018surface}.
However, because the measurements are generally carried out under loading conditions, it is difficult to extract the contribution of dispersion force from the various forces that make up the surface force.

AFM can accurately measure the interaction force between two microparticles \citep{Butt:2005force}. \citet{Mohideen:1998precision} first precisely measured the dispersion force between a metal sphere and plate with AFM.
An accurate measurement involves strictly excluding the contributions of forces other than the dispersion force, which is typically very difficult \citep{Rodriguez:2011casimir}. Therefore, at present, accurate measurements are limited to a few special geometries (sphere and plate) \citep{Bressi:2002measurement, Garrett:2018measurement}.
Although some studies have investigated particles with complex morphologies \citep{Moutinho:2017adhesion, Zhao:2020correlation} with AFM, the environment was not strictly controlled, and the adhesion or cohesion was measured instead of the dispersion force.
In addition, in the field of chemical engineering, the centrifugal method has been used to study the effects of particle morphology on cohesion and adhesion \citep{Stevenson:2021effects, Nagaashi:2021cohesion}. However, the centrifugal method is a rougher technique than SFA and AFM. Additionally, cohesion and adhesion were measured in this study as opposed to the dispersion force.
Recently, in microelectromechanical systems (MEMS), special equipment has been developed to quantify the effect of the boundary geometry on the dispersion force \citep{Tang:2017measurement, Wang:2021strong, Sedighi:2016casimir, Svetovoy:2020measuring, Soldatenkov:2022dispersion}. However, the research objects in these studies were artificial plates with relatively regular surfaces (rectangular silicon gratings and corrugated plates) rather than real particles with complex morphologies.
Thus, in summary, previous experimental studies mainly focused on how particle geometry influences adhesion or cohesion and did not precisely determine the effect of the dispersion force.
Critical experiments on the dispersion force are limited to a few special geometries and materials.
Therefore, a general law about the influence of geometry on the dispersion force has not yet been established due to the lack of experimental data on real particles with complex morphologies.

Other effective methods include numerical calculations.
Thus far, two approximate algorithms have been commonly applied because of their simplicity: pairwise summation approximation (PWS) and proximity force approximation (PFA) \citep{Parsegian:2005van,Sonnenberg:2005numerical,Hopkins:2015disentangling}.
PWS, which is also known as Hamaker summation \citep{Hamaker:1937london}, calculates the dispersion force between two macroscopic bodies according to the pairwise summation of volumetric elements interacting through the vdW-Casimir force, which can be established based on dipolar dispersive interactions.
PFA, which is also known as the Derjaguin approximation, models the interaction between nearby surfaces as additive line-of-sight interactions between infinitesimal, planar surface elements (computed via the Lifshitz formula \citep{Dzyaloshinskii:1961general}).
Both algorithms are based on the assumption of additivity, with the force calculated by simply summing the force contributions of the surface or dipole interactions.
Unfortunately, a concise and convenient assumption always comes at the expense of accuracy.
In fact, the interaction between two dipoles/surfaces is influenced by nearby dipoles/surfaces, which is ignored in the additivity assumption.
Moreover, the multiple scattering caused by multibody interactions cannot be ignored for condensed matter.
The discrepancy between additive approximations and exact calculations is often referred to as nonadditivity.
A detailed explanation of nonadditivity can be found in the review by \citet{Rodriguez:2011casimir}.
Nonadditivity considerably reduces the accuracy of PWS and PFA techniques for handling condensed matter with complex morphology, which has been confirmed by several theoretical and experimental studies \citep{Emig:2001probing, Gies:2006casimir, Bitbol:2013pairwise, Venkataram:2016nonadditivity, Wang:2021strong}.
Therefore, these two additivity methods cannot be applied to study the influence of geometry on the dispersion forces of real particles.

Obviously, the interactions among each dipole are impossible to calculate except at the subnanometer scale \citep{Venkataram:2016nonadditivity, Venkataram:2017unifying}.
As previously mentioned, dispersion forces originate from quantum fluctuations. In terms of field fluctuations, the issues caused by nonadditivity can be addressed.
This kind of algorithm originated in the 1950s \citep{Dzyaloshinskii:1961general} and was designed to work for one special geometry.
Since 2007 \citep{Emig:2007casimir}, dramatic progress has been made in this field, and general-purpose schemes for arbitrary materials and various geometric configurations have been developed \citep{Reid:2013fluctuating}.
These schemes can be roughly divided into two categories according to the choice of physical basis: fully quantum mechanical approaches and semiclassical approaches.
The former include the path integral (or scattering) approach \citep{Emig:2007casimir,Emig:2009orientation} and the lattice field approach \citep{Chernodub:2016casimir, Chernodub:2020casimir}, which are highly efficient in certain geometries but not flexible enough for general geometries, especially geometries with sharp corners \citep{Reid:2013fluctuating}.
In the semiclassical approach (also known as the stress tensor approach) \citep{Rodriguez:2007virtual, Rodriguez:2009casimir}, the computation of the fluctuation force can be reduced to solving the classical electromagnetic scattering problem according to the fluctuation-dissipation theorem (FDT).
Thus, many classical electrodynamics methods can be applied after a few important modifications (detailed in reviews \citet{Rodriguez:2011casimir} and \citet{Johnson:2011numerical}).
In addition to these two schemes, the fluctuating surface current (FSC) technique (which will be introduced in detail in Section $\mathsection$\ref{sec:fsc}) developed by \citet{Reid:2009efficient} can be regarded as a third scheme that combines the advantages of the previous two approaches.
In these nonadditivity algorithms, interactions among instantaneous dipoles are not considered and are replaced by the global optical response of the material, which can be described by macroscopic dielectric functions.
Based on the dielectric functions of the materials, the dispersion force can be accurately calculated by considering the complex electromagnetic modes and surface scattering properties \citep{Parsegian:2005van}.
The accuracy of these algorithms has been confirmed by strict experiments, even in cases with complex geometric configurations \citep{Reid:2013fluctuating, Tang:2017measurement, Wang:2021strong}.
As \citet{Svetovoy:2015influence} noted, these nonadditivity algorithms should be the most suitable numerical methods for studying the influence of particle geometry on dispersion forces.
The generality of these methods does, however, come at a price, with even the most sophisticated of formulations requiring thousands or hundreds of thousands of scattering calculations to be performed.
Moreover, the calculation accuracy needs to be guaranteed by obtaining accurate dielectric functions for the materials, which are usually empirical.
Therefore, at present, these algorithms are still limited to a small number of special materials and geometries, and few related calculations have been reported for real particles with complex morphologies.

In summary, no general law about the influence of particle geometry on dispersion forces has been established. Most previous experimental studies focused on surface  or cohesion forces instead of the dispersion force, which requires that the experimental environment be strictly controlled, and critical experimental studies were limited to a few special geometries and materials. Moreover, the conventional calculation methods are based on the assumption of additivity, which is not suitable for real particles with complex morphologies. Recent nonadditivity algorithms were developed based on field fluctuations and can be applied to accurately calculate dispersion forces in complex geometric configurations; however, these techniques are also limited to a small number of geometries and materials.
In this research, we apply a kind of this exact technique to perform large-scale brute-force computations to develop a general law on how particle geometry influences dispersion forces.
The remainder of this paper is organized as follows.
First, we introduce a reasonable characterization of complex morphologies based on spherical empirical modal decomposition (SEMD), including a multiscale decomposition of the morphology and a three-level characterization of real particles. Then, an accurate numerical model for determining the dispersion force beyond nonadditivity, namely, the FSC algorithm, is briefly introduced.
Then, the dispersion forces of particles with different morphologies are calculated. We produce different geometric configurations for the same morphology through random rotations.
According to the calculation results, we present a general law on the influence of multiscale surface fluctuations on the dispersion force. Furthermore, through multistage approximations, we propose a convenient formula for predicting the dispersion force between real complex shaped particles that can be used in engineering applications.

\section{Methods}
\subsection{Characterization of particle geometry}
\label{sec:geoc}
As previously mentioned, real microparticles (which are encountered in daily life, such as pollen, sand, dust, and flour) have a variety of irregular shapes that differ substantially from any idealized geometry, regardless of whether the particles are natural or artificial. Ideal geometries such as spheres and cubes can be fully characterized by simple parameters such as the radius, edge length, area and volume. However, these simple parameters cannot be applied to characterize the geometrical properties of complex irregular particles. In fact, despite almost a century of discussion, a reasonable characterization of irregular geometries has not yet been established \citep{Feng:2020three, Angelidakis:2021shape}.

There is a consensus that irregular geometries are too complex to be characterized by a single parameter at one scale \citep{ISO:2008, kamarainen:2018harmonic}.
The existing particle morphology characterization methods can be roughly divided into two categories: Euclidean descriptors and frequency domain descriptors.
The Euclidean descriptor is a traditional descriptor that extracts description parameters by combining basic parameters such as the axial length, volume, surface area, and local curvature of the particles.
Dozens of Euclidean descriptors have been proposed since the first was established in \citet{Wadell:1932volume}.
The frequency domain descriptor is a recent descriptor \citep{Luerkens:1982morphological, Zhao:20163d} that decomposes the morphology into a combination of orthogonal functions by appropriate function transformations (commonly, Fourier transforms for 2D morphologies and spherical harmonic transforms for 3D morphologies) and extracts different order coefficients of the transformation functions to characterize particles at different scales. These descriptors reflect the multiscale characteristics of the morphology.

\begin{figure}
\centering
\includegraphics[width=1.0\textwidth]{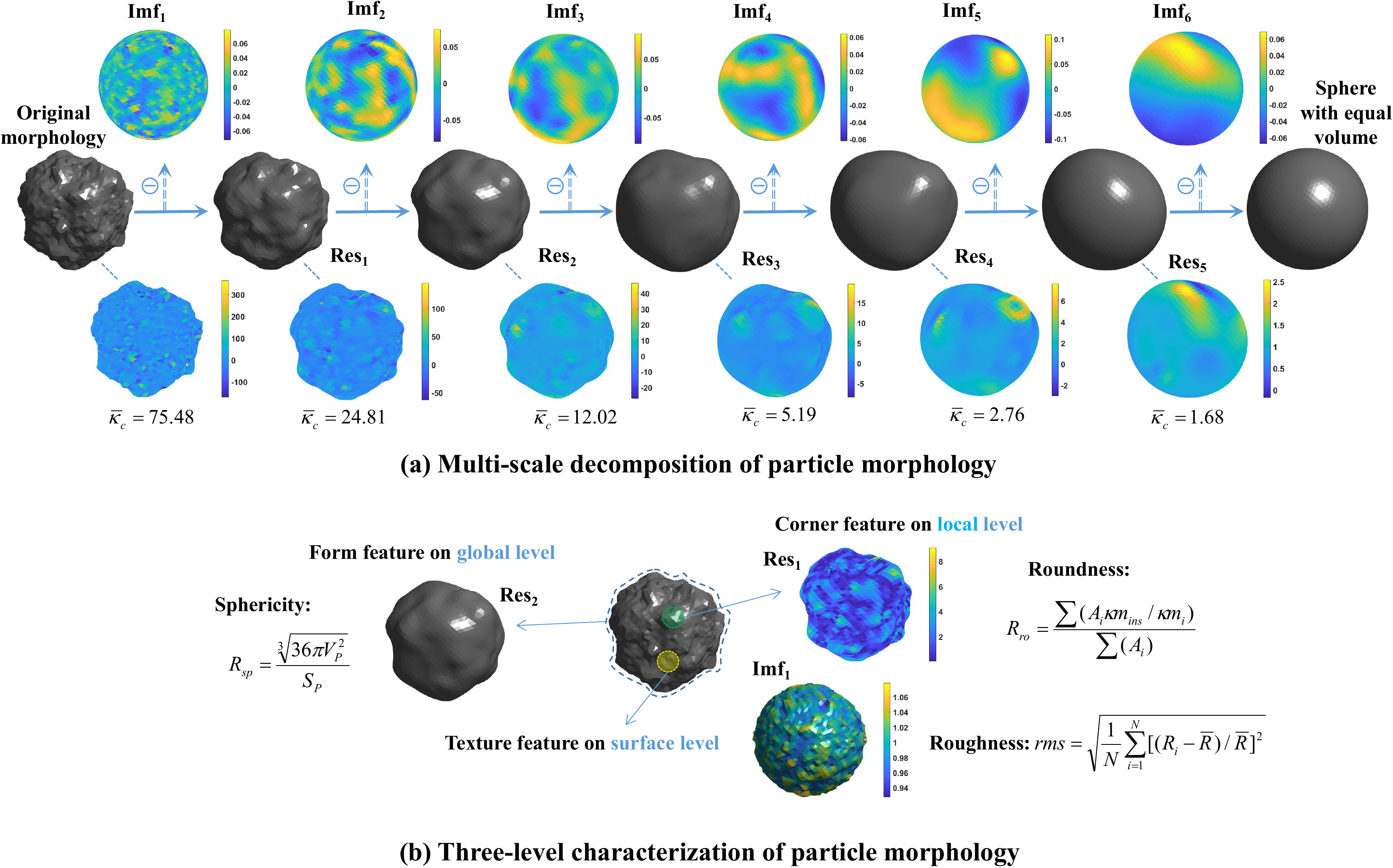}
\caption{Characterization of particle geometry. (a) is an example of scale decomposition using SEMD. The top row shows the intrinsic mode function ($Imf_i$) corresponding to each order of decomposition, i.e., the part that is adaptively filtered out; the middle row shows the residual shape ($Res_i$) after filtering; and the last row shows the Gaussian curvature (see \citet{Schroder:2013minkowski} for detailed definition) distribution of the residual shape, and the average of the Gaussian curvatures of the surface maximums ($\overline{\kappa}_{c}$) was taken as the scale characterization parameter. (b) is the schematic of the three descriptors: sphericity for the global form feature, roundness for the local corner feature and roughness for the surface texture feature. See \citet{Zheng:2015traditional} and \textcolor{blue}{supplementary materials} for the detailed definition of the descriptors.}
\label{fig:1geocha}
\end{figure}

Since the two types of descriptors provide different views on morphological characterization, we employ slightly modified versions of both descriptors in this study.
Here, instead of the frequency domain descriptors themselves, we consider the decomposition (function transformation) process in this study, which decomposes the complex morphologies of real particles into surface fluctuations at different scales.
The commonly used Fourier transform and spherical harmonic transform are suitable only for stationary signals because they are global transformations in the time domain, while the morphology of real particles are typically nonstationary signals (with nonstationary characteristics such as sharp corners, planes, edges and other abrupt features).
Therefore, in the decomposition process, we introduce the spherical empirical mode decomposition (SEMD) technique \citep{Fauchereau:2008empirical,Rehman:2010multivariate}, the core algorithm of Hilbert-Huang transform \citep{Huang:1998empirical}, which works well for nonstationary signals.
The physical essence of SEMD is a data-adaptive spatial filter. As shown in Fig. \ref{fig:1geocha}(a), taking the average of the Gaussian curvatures of the surface maximums ($\overline{\kappa}_{c}$) as the scale characterization parameter, the fine details (fluctuations with larger curvatures) are filtered out in turn along the decomposition process adaptively.
The detail of the SEMD algorithm were described in the \textcolor{blue}{supplementary materials}
and the corresponding program can be found in \url{https://github.com/yfliu088/GMAP}.
Based on this decomposition, the influence of multiscale surface fluctuations on the dispersion force can be studied (Sec. $\mathsection$\ref{sec:multi}).

For the traditional Euclidean descriptors, we draw on recent researches in the field of geotechnical engineering on the morphological characterization of soil particles \citep{Zheng:2015traditional, Zhou:2018three}.
As shown in Fig. \ref{fig:1geocha}(b), 3 descriptors at 3 observation levels were selected to comprehensively characterize the particle morphology, including sphericity for the global level, roundness for the local level and roughness for the surface level.
Here, to prevent interactions among the different observation levels, we first filtered the surface using the aforementioned SEMD before calculating the values of each descriptor (details in \textcolor{blue}{supplementary materials}).
In Sec. $\mathsection$\ref{sec:3levelc}, we establish relationships between the dispersion forces and these common descriptors by performing a large number of numerical calculations on particles with different morphologies.

\subsection{Exact numerical model for the dispersion force beyond the additivity assumption}
\label{sec:fsc}
To study the influence of particle geometry on dispersion forces, we need to handle many particles with different complex morphologies, which requires a sufficiently flexible computational framework that can handle various morphologies. The FSC technique, a unified formalism capable of handling arbitrary materials in various geometries \citep{Reid:2011fluctuating, Reid:2015efficient}, is the best nonadditivity algorithm for investigating this problem.

\begin{figure}
\centering
\includegraphics[width=0.6\textwidth]{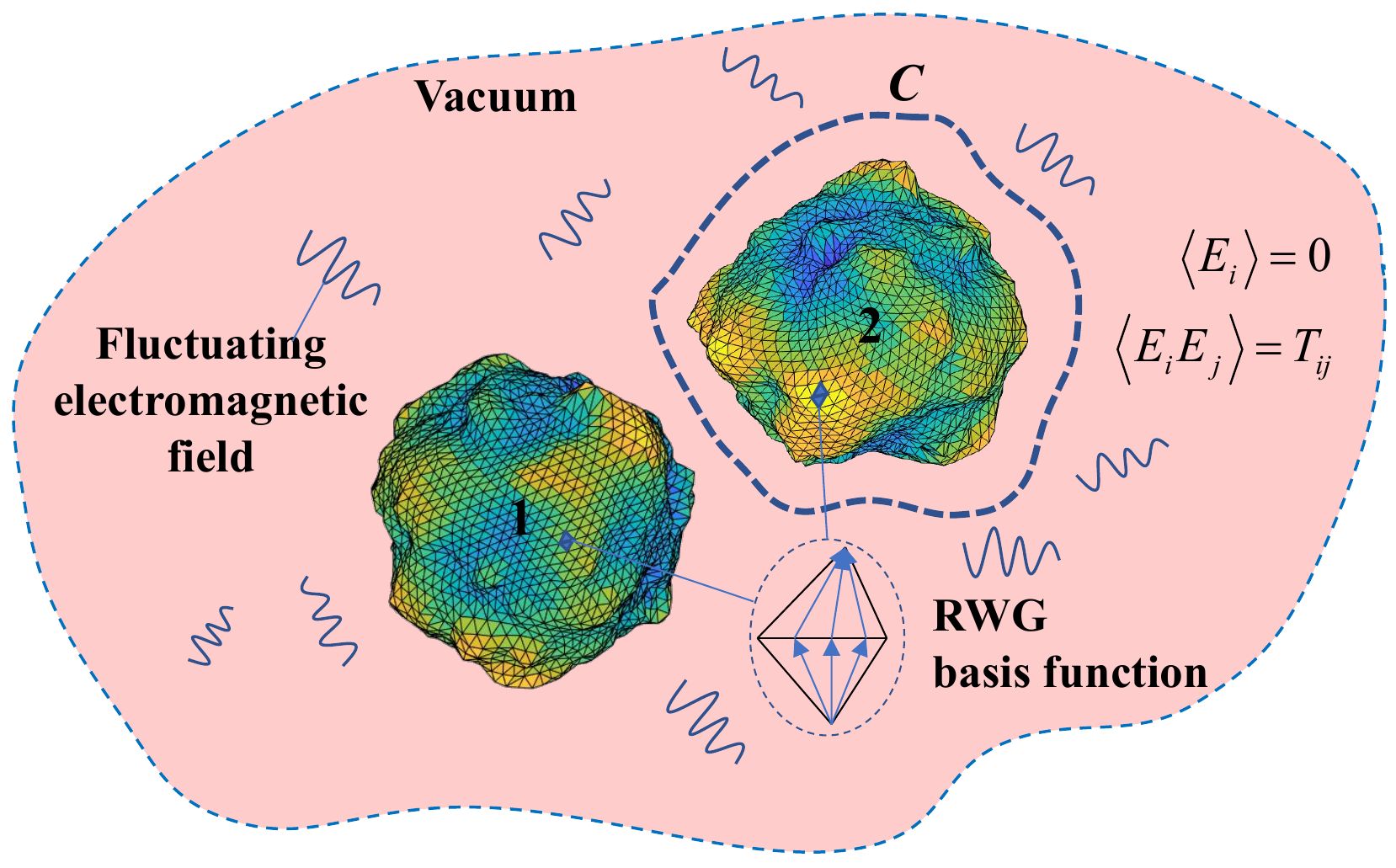}
\caption{Schematic depiction of the FSC algorithm.}
\label{fig:schfsc}
\end{figure}

The FSC algorithm is a well-established algorithm, and the critical elements of this technique are introduced below.
As shown in Fig. \ref{fig:schfsc}, two neutral compact objects placed in a vacuum derive their dispersion force according to the surrounding fluctuating electromagnetic field.
In the absence of any external force field, the average value of any single field component vanishes; however, the average values of the off-diagonal products of the field components are generally nonzero and may be related to a force density (the fluctuation-averaged Maxwell stress tensor ($T_{ij}$)) that we can use to compute the dispersion forces.

As shown in Fig. \ref{fig:schfsc}, the $i$-directed dispersion force on compact object $2$ can be determined by integrating the fluctuation-averaged stress tensor over any closed bounding surface $C$ surrounding this body:
\begin{eqnarray}
\mathcal{F}_{i}=\int_{0}^{\infty}\frac{d\xi}{\pi}F_{i}(\xi),
\label{eq:fsci1}
\end{eqnarray}
\begin{eqnarray}
F_{i}(\xi)=\oint_{\mathcal{C}} \left \langle T_{ij}(\boldsymbol{x},\xi) \right \rangle\hat{\boldsymbol{n}_j}(x) d\boldsymbol{x},
\label{eq:fsci2}
\end{eqnarray}
here, Eq. \eqref{eq:fsci1} indicates that the dispersion force can be obtained by integrating the contributions of all imaginary frequencies $\xi=i\omega$. In Eq. \eqref{eq:fsci2}, $\hat{\boldsymbol{n}_j}(x)$ is the inward-directed unit normal to $C$ at $x$, and the expectation value $\left \langle T_{ij} \right \rangle$ can be written in terms of the components of the electric and magnetic fields:
\begin{equation}
\begin{aligned}
T_{ij}(\boldsymbol{x},\xi)=&\varepsilon(\boldsymbol{x},\xi)\left[\left \langle E_{i}(\boldsymbol{x})E_{j}(\boldsymbol{x}) \right \rangle_{\xi}-\frac{\delta_{ij}}{2}\sum_{k}\left \langle E_{k}(\boldsymbol{x})E_{k}(\boldsymbol{x}) \right \rangle_{\xi} \right]\\
&+\mu(\boldsymbol{x},\xi)\left[\left \langle H_{i}(\boldsymbol{x})H_{j}(\boldsymbol{x}) \right \rangle_{\xi}-\frac{\delta_{ij}}{2}\sum_{k}\left \langle H_{k}(\boldsymbol{x})H_{k}(\boldsymbol{x}) \right \rangle_{\xi} \right].
\end{aligned}
\label{eq:fsci3}
\end{equation}
Thus far, the calculation of the dispersion force has been reduced to calculating the fluctuation-averaged products of the field components at each imaginary frequency.

Crucial work by Lifshitz et al. in the 1950s \citep{Dzyaloshinskii:1961general,Lifshitz:2013statistical} established the relationship between the fluctuation-averaged products of the field components and the scattering portions of the dyadic Green's functions in classical electromagnetism based on the fluctuation-dissipation theorem:
\begin{equation}
\begin{aligned}
&\left \langle E_{i}(\boldsymbol{x})E_{j}(\boldsymbol{x}) \right \rangle_{\xi}=\hbar\xi\mathcal{G}_{ij}^{\text{EE}}(\xi;\boldsymbol{x},\boldsymbol{x'}), \\
&\left \langle H_{i}(\boldsymbol{x})H_{j}(\boldsymbol{x}) \right \rangle_{\xi}=\hbar\xi\mathcal{G}_{ij}^{\text{MM}}(\xi;\boldsymbol{x},\boldsymbol{x'}).
\end{aligned}
\label{eq:fsci4}
\end{equation}
Here, $\mathcal{G}_{ij}^{\text{EE}}(\xi;\boldsymbol{x},\boldsymbol{x'})$ and $\mathcal{G}_{ij}^{\text{MM}}(\xi;\boldsymbol{x},\boldsymbol{x'})$ are the scattered portions of the electric and magnetic fields at $x$ due to the current sources at $x'$, respectively.
According to Eq. \eqref{eq:fsci4}, the calculation of the dispersion force can be completely reduced to solving the classical electromagnetic scattering problem.
Therefore, various classical methods can be exploited to calculate the dispersion forces originating from quantum fluctuations, with a few important modifications. The key modification is the use of the so-called Wick rotation ($\xi=i\omega$) to stabilize the calculation, which ensures that all frequency-dependent quantities ($\varepsilon$, $\mu$, $\left \langle\right \rangle$, $\mathcal{G}$) in Eqs. \eqref{eq:fsci1}-\eqref{eq:fsci4} are considered in the imaginary frequency.

For scattering problems with homogeneous geometries embedded in homogeneous media, the boundary element method (BEM) can be applied to effectively use the known solutions of Maxwell's equations, thus greatly improving the computational efficiency. Therefore, the BEM rather than other computational electromagnetic methods (such as the finite difference or finite element methods) was used in the FSC algorithm. Taking perfectly electrically conducting (PEC) bodies embedded in a vacuum as an example, the scattering part of the dyadic Green's function $\mathcal{G}$ can be obtained with the BEM as:
\begin{equation}
\begin{aligned}
&\mathcal{G}_{ij}^{\text{EE}}(\xi;\boldsymbol{x},\boldsymbol{x'})=-\sum_{\alpha\beta} \left\langle\Gamma_{i}^{\text{EE}(\boldsymbol{x})} \big| f_{\alpha}\right\rangle \left[M^{-1} \right]_{\alpha\beta}\left\langle\Gamma_{j}^{\text{EE}(\boldsymbol{x})} \big| f_{\beta}\right\rangle,  \\
&\mathcal{G}_{ij}^{\text{MM}}(\xi;\boldsymbol{x},\boldsymbol{x'})=-\sum_{\alpha\beta} \left\langle\Gamma_{i}^{\text{ME}(\boldsymbol{x})} \big| f_{\alpha}\right\rangle \left[M^{-1} \right]_{\alpha\beta}\left\langle\Gamma_{j}^{\text{EM}(\boldsymbol{x})} \big| f_{\beta}\right\rangle.
\end{aligned}
\label{eq:fsci5}
\end{equation}
Here, $\Gamma$ denotes the vacuum dyadic Green's functions, and $\{f_{\alpha}\}$ refers to a set of localized tangential vector-valued basis functions depending on the surface discretization (i.e., the boundary elements). $\left\langle\Gamma\big| f_{\alpha}\right\rangle$ refers to the interactions between $\Gamma$ and $f_{\alpha}$, namely, $\int_{\text{sup}\boldsymbol{f}_\alpha}\Gamma\cdot f_{\alpha} d\boldsymbol{x}$, and $M$ is the BEM matrix, which describes the interactions among the basis functions in the exterior medium as follows: $M_{\alpha\beta}=\int_{\text{sup}\boldsymbol{f}_\alpha}\int_{\text{sup}\boldsymbol{f}_\beta}f_{\alpha}(\boldsymbol{x})\cdot\Gamma^{\text{EE}}\cdot f_{\beta}(\boldsymbol{x}')d\boldsymbol{x}d\boldsymbol{x}'$.

Eq. \eqref{eq:fsci5} can be used to calculate $\mathcal{G}$ at each point on the closed bounding surface $\mathcal{C}$; thus, the dispersion force can be obtained via integration.
However, \citet{Reid:2015efficient} carried out a further analytical derivation based on Eq. \eqref{eq:fsci5}, simplifying the calculation of the dispersion force as a compact matrix-trace formula.

Substituting Eq. \eqref{eq:fsci5} into Eq. \eqref{eq:fsci2}, according to the symmetry of $M$, the dispersion force at a certain frequency can be written as follows:
\begin{eqnarray}
F_{i}(\xi)=\frac{\hbar}{2}Z_0\kappa\sum_{\alpha\beta}\left[M^{-1}\right]_{\alpha\beta}\int_{\text{sup}\boldsymbol{f}_\alpha}d\boldsymbol{r}\int_{\text{sup}\boldsymbol{f}_\beta}d\boldsymbol{r'}f_{\alpha k}(\boldsymbol{r})\cdot\bar{\mathcal{L}}_{ikl}(\boldsymbol{r},\boldsymbol{r'})\cdot f_{\beta l}(\boldsymbol{r'}),
\label{eq:fsci6}
\end{eqnarray}
Here, $\bar{\mathcal{L}}_{ikl}(\boldsymbol{r},\boldsymbol{r'})$ is a symmetrized version of the integral kernel along the closed bounding surface $\mathcal{C}$, which has the following special property:
\begin{eqnarray}
\bar{\mathcal{L}}_{ikl}(\boldsymbol{r},\boldsymbol{r'})=\left\{
\begin{array}{lr}
    0 &\text{if } \boldsymbol{r},\boldsymbol{r'} \text{ lie on the same side of } \mathcal{C}\\
     \frac{\partial}{\partial}G_{kl}(\boldsymbol{r}^i-\boldsymbol{r}^e) &\text{if } \boldsymbol{r},\boldsymbol{r'} \text{ lie on opposite side of } \mathcal{C}
\end{array}
\right.
\label{eq:fsci7}
\end{eqnarray}
where $G$ refers to the photon Green's function and $\boldsymbol{r}^i(\boldsymbol{r}^e)$ indicates whether $\boldsymbol{r}$ and $\boldsymbol{r}'$ lie in the interior (exterior) of $\mathcal{C}$. According to Eq. \eqref{eq:fsci7}, for the case shown in Fig. \ref{fig:schfsc}, the integral kernel $\bar{\mathcal{L}}$ is nonzero only when $f_\alpha$ and $f_\beta$ are on the surfaces of particles $1$ and $2$, respectively.
Based on this special property, Eq. \eqref{eq:fsci6} can be reduced to the following matrix-trace form:
\begin{eqnarray}
F_{i}(\xi)=-\frac{\hbar}{2}\text{Tr}\left\{\boldsymbol{M}^{-1}\cdot\frac{\partial \boldsymbol{M}}{\partial\boldsymbol{r}_i}\right\}.
\label{eq:fsci8}
\end{eqnarray}
Then, the full dispersion force in Eq. \eqref{eq:fsci1} becomes
\begin{eqnarray}
\mathcal{F}_{i}=-\frac{\hbar}{2}\int_{0}^{\infty}d\xi\text{Tr}\left\{\boldsymbol{M}^{-1}\cdot\frac{\partial \boldsymbol{M}}{\partial\boldsymbol{r}_i}\right\}.
\label{eq:fsci9}
\end{eqnarray}
At this point, the calculation of the dispersion force is finally reduced to calculate the BEM matrix $M$. For general materials, the derivation is similar, and the final formulae have the form shown in Eq. \eqref{eq:fsci9}, except that $M$ needs to be replaced by the PMCHW matrix.

The solution of the BEM matrix requires a reasonable surface discretization and the corresponding localized basis functions. As shown in Fig. \ref{fig:schfsc}, the surfaces of compact objects can be discretized by Delaunay triangulation, and the RWG basis functions (named after \citet{Rao:1982electromagnetic}) based on two adjacent triangular planes are applied in the FSC algorithm. For the present study, this choice has the additional benefit of sharing the same set of meshes with the aforementioned morphological analysis.

The above description allows the dispersion force to be computed at zero temperature $T=0^+$. In the case of a nonzero temperature ($T>0$), the description can be modified by converting the integral in Eq. \eqref{eq:fsci9} into a sum over a series of discrete imaginary frequencies:
\begin{eqnarray}
\mathcal{F}_{i}=\frac{2\pi k_{B}T}{\hbar}\sideset{}{'}\sum_{n=0}^{\infty}F_{i}(\xi_n), \text{     }\xi_n=n\frac{2\pi k_{B}T}{\hbar}
\label{eq:fsci10}
\end{eqnarray}
where $k_B$ is Boltzmann's constant and $\sideset{}{'}\sum$ indicates a sum with weight $1/2$ for the $n=0$ term. The frequencies $\xi_n=n\frac{2\pi k_{B}T}{\hbar}$ are known as Matsubara frequencies.

The accurate computation of the dispersion force using the FSC technique requires that accurate dielectric data of the materials are available.
As shown in Eq. \ref{eq:fsci10}, theoretically, the contributions of different Matsubara frequencies to the full band must be calculated.
For interbody distances at the micro-nanometer scale, frequencies in the ultraviolet (UV) region have a considerable impact on the dispersion force \citep{Parsegian:2005van}.
However, it is not possible to obtain tabulated dielectric data for all dispersive materials in the full band.
Reviewing the available dielectric data, data in the infrared and visible regions are abundant, while data in the UV region are lacking.
This data imbalance occurs because few techniques exist for measuring optical constants in the UV region \citep{Moazzami:2021self}.
Therefore, different ideal dielectric function models have been introduced to predict the frequency without measurement data, such as the Drude model, plasma model, Lorentz model and Ninham-Parsegian model, which have been commonly used in recent studies.
In addition to these models, a novel empirical modified harmonic oscillator model was recently introduced by \citet{Moazzami:2021self}.
In this model, dielectric data of 55 common materials obtained from different sources were compiled and evaluated according to optical sum rules and Kramers-Kronig relations to ensure internal consistency. On the basis of these data and the optical bandgap, density, and chemical composition of the materials, the model parameters can be accurately determined. The accuracy of this dielectric function model was validated by the experimentally measured dispersion force in a planar configuration. The general form of the model can be expressed as follows:
\begin{eqnarray}
\varepsilon(\xi_n)=\left\{
\begin{array}{lr}
    \sum\limits_{j}\frac{C_j}{1+\left(\frac{\xi}{\omega_j}\right)^{\alpha_j}}, \text{     }  (j\le 3) &\text{for semiconductors and insulators}\\
     \frac{C_{UV}}{1+\left(\frac{\xi}{\omega_{UV}}\right)^{\alpha_{UV}}}+\frac{\omega_p^2}{\xi^2+\gamma\xi} &\text{for metals}
\end{array}
\right.
\label{eq:fsci11}
\end{eqnarray}
Here, we select four materials with different properties, namely, \ch{Au} (metal), \ch{Si} (semiconductor), \ch{SiO_2} (inorganic insulator) and \ch{C_2H_4} (organic insulator), and calculate the dispersion force under complex morphology. The specific details of the dielectric functions can be found in the original literature of \citet{Moazzami:2021self}.

After addressing the nonadditivity and dielectric functions, the only remaining factor that may affect the accuracy is the discrete error in the numerical calculation. In the \ref{sec:appA}, we provide a detailed discretization study under different geometric configurations. Considering the efficiency, the number of surface triangles was set to 5000, and the accuracy can be guaranteed when the surface curvature $\overline{\kappa}_{c}<100$.

\subsection{Geometric configurations in the computational implementation}
\label{sec:conf}
To study the influence of particle geometry on dispersion forces, data of real particles with varying morphologies are first needed. To generalize the calculations and obtain sufficient ranges for each characterization parameter, a numerical method proposed by \citet{Wei:2018simple} for randomly constructing real particles with different morphologies was employed.

\begin{figure}
\centering
\includegraphics[width=0.6\textwidth]{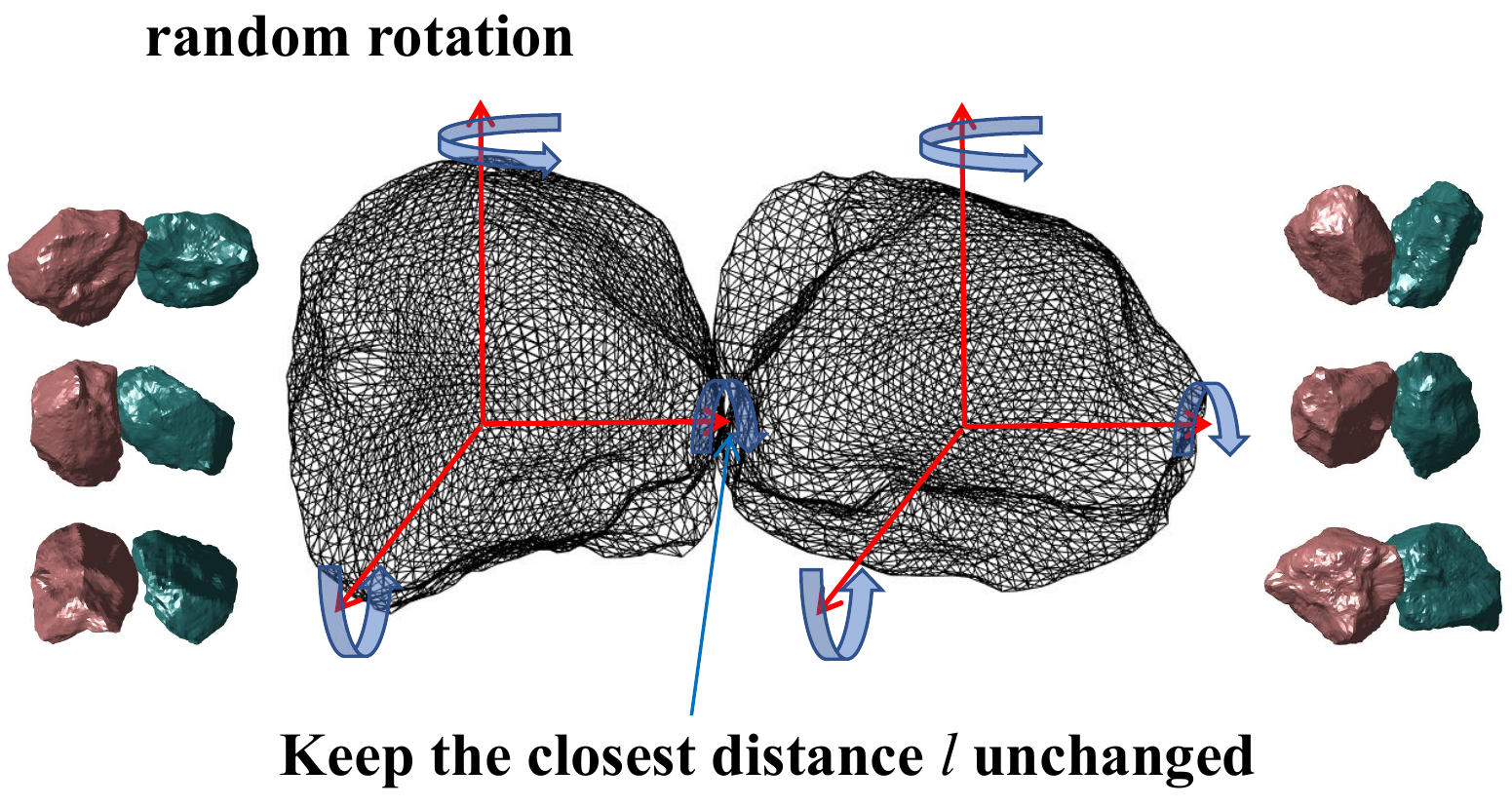}
\caption{Geometric configurations in the computational implementation for the arbitrary particles. The middle shows the random rotation of two particles along three axes while keeping the closest distance between the two particles constant. The two sides show examples after random rotation.}
\label{fig:3geoconf}
\end{figure}

Unlike spheres, arbitrary shapes are anisotropic, resulting in two problems: the definition of the distance between particles is uncertain, and the geometric configuration changes as the particles rotate.
In practical applications, we are concerned with the mean and standard deviation of the dispersion forces.
As shown in Fig. \ref{fig:3geoconf}, we maintain the closest distance between the particles, which we define as the distance between the particles, and calculate the mean and standard deviation of the dispersion force by randomly rotating the particles to produce different configurations.

\section{Results and Discussion}
\subsection{Comparison between additivity and nonadditivity methods}
\begin{figure}
\centering
\includegraphics[width=1.0\textwidth]{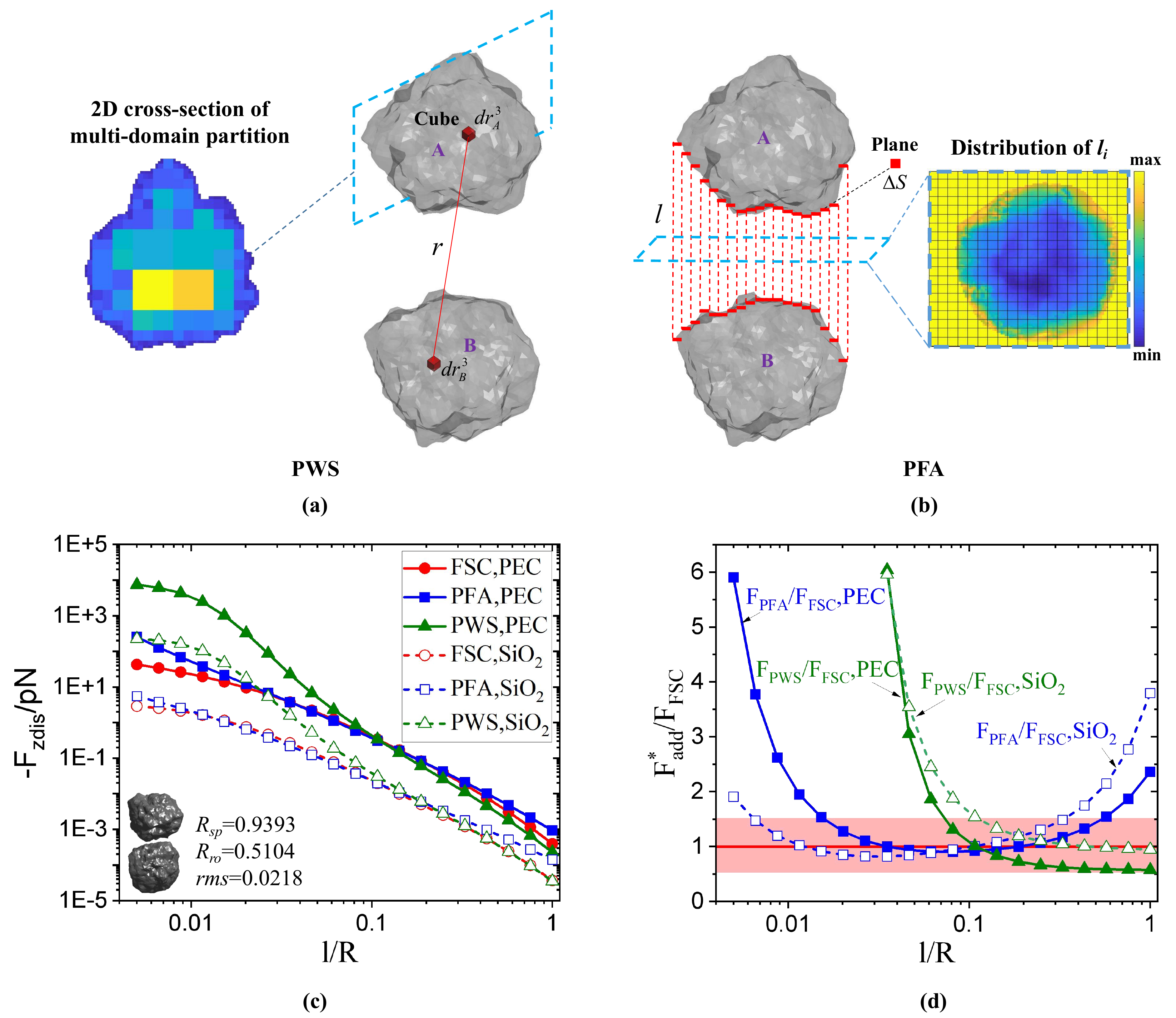}
\caption{Comparison between additivity and nonadditivity methods. (a) is the schematic of the PWS algorithm, and the left 2D cross-section shows the proposed multi-domain partition approach. (b) is the schematic of the PFA algorithm, and the right contour shows the distribution of the distance of each plane-plane configuration of the two curved surfaces. (c) and (d) show the comparison of the computed results of the three algorithms, where (c) shows the calculated dispersion forces and (d) shows the results of normalizing the two additive algorithms using FSC.}
\label{fig:pfapws}
\end{figure}

For regular geometries, previous literature has extensively compared additivity and nonadditivity methods. Here, for particles with complex shapes, we first present a quantitative comparison between additivity and nonadditivity methods.
Schematic diagrams of two commonly used algorithms (PWS and PFA) based on the additivity assumption are shown in Fig. \ref{fig:pfapws}(a).
For the PWS technique, we use a modified version of the Hamaker summation approximation by considering Lifshitz theory and the Clausius-Mossotti relation \citep{Wang:2021strong}. In this algorithm, the potential energy originating from the dispersion force between particles $A$ and $B$ can be expressed as:
\begin{equation}
\begin{aligned}
&G_{AB}(T)_{PWS}=-\frac{\hbar}{\pi}\left(\frac{3}{4\pi}\right)^2\int_{V_A}d^3_{r_A}\int_{V_B}d^3_{r_B}\sideset{}{'}\sum_{n=0}^{\infty}g(\xi_n),\\
&g(\xi_n)=\left(\frac{\varepsilon(i\xi_n)-1}{\varepsilon(i\xi_n)+2}\right)^2\left[\frac{3}{r^6}+\left(\frac{\xi_n}{c}\right)\frac{6}{r^5}+\left(\frac{\xi_n}{c}\right)^2\frac{5}{r^4}+\left(\frac{\xi_n}{c}\right)^3\frac{3}{r^3}+\left(\frac{\xi_n}{c}\right)^4\frac{1}{r^2}\right]e^{-\frac{2\xi_n r}{c}}.
\end{aligned}
\label{eq:pws}
\end{equation}
$\{\xi_n\}$ are the Matsubara frequencies at temperature $T$, as shown in Eq. \eqref{eq:fsci10}, and $r$ refers to the distance between the two cubes $d^3_{r_A}$ and $d^3_{r_B}$. The dispersion force is obtained by taking the spatial derivative of $G_{AB}(T)_{PWS}$. This calculation takes into account the dispersion property, retard effect and temperature correction.
For the PFA algorithm, the potential energy per unit area in the plane-plane configuration must be determined first, the exact solution of which was derived by \citet{Lifshitz:1956theory}:
\begin{eqnarray}
G_{pp}(T,l)=\frac{k_BT}{2\pi}\sideset{}{'}\sum_{n=0}^{\infty}\int_0^{\infty}k_{\bot}dk_{\bot}\times\sum_{\alpha=\text{TE},\text{TM}}\ln[1-r_{\alpha}^2(i\xi_n,k_{\bot})e^{-2lq_n}],
\label{eq:pfa1}
\end{eqnarray}
where $k_{\bot}$ denotes the magnitude of the projection of the wave vector onto the plane of the plates, $r_{\alpha}(i\xi_n,k_{\bot})$ denotes the Fresnel reflection coefficient for the two independent polarizations $\alpha=\text{TE, TM}$ of the electromagnetic field, $l$ refers to the distance between the two planes and $q_n=\sqrt{\xi_n^2/c^2+k_{\bot}^2}$.
The dispersion property, retard effect and temperature correction are also considered in this analytical solution. Then, the total potential energy between the curved surfaces is obtained by summing the potential energy of each plane-plane configuration:
\begin{eqnarray}
G_{AB}(T)_{PFA}=\sum G_{pp}(T,l_i)\Delta S_{i},
\label{eq:pfa2}
\end{eqnarray}
where $l_i$ and $\Delta S_{i}$ denote the distance and surface area of each plane-plane configuration of the curved surfaces, respectively, as shown in Fig. \ref{fig:pfapws}(a). Finally, the dispersion force is obtained by taking
the spatial derivative of $G_{AB}(T)_{PFA}$.

In this comparative study, the optimal versions of both additivity algorithms are used, taking into account corrections for factors other than geometry. Moreover, the computational complexity of the PWS algorithm is $O(L^6)$ ($L$ refers to the linear resolution of the volume integral). However, for particles with complex shapes, the computational complexity becomes unacceptable if the resolution is large enough to capture boundary variations. Thus, we propose a multidomain partitioning approach, as shown in Fig. \ref{fig:pfapws}(a). With this approach, the computational complexity is significantly reduced to an acceptable level. The optimized programs for the two additivity algorithms are provided at \url{https://github.com/yfliu088/vdw-casimir}.

A complex-shaped particle was selected, the materials were set to a perfectly electrically conducting (PEC) metal and a dispersive medium \ch{SiO_2}, and the temperature was set to room temperature ($298.15 K$). A comparison of the three algorithms as the distance changes is shown in Fig. \ref{fig:pfapws}(c) and (d).
Overall, compared to the exact nonadditivity algorithm, the additivity algorithms overestimate the dispersion force.
For complex-shaped particles, the PFA technique performs better than the PWS approach. The relative errors in the PFA and PWS methods can reach up to $\sim10$ times and $\sim1000$ times, respectively.
For each additivity algorithm, the error trends of the PEC and dispersive medium are consistent. For the PFA technique, the error at intermediate distances is relatively small, while for the PWS method, the error is smaller at longer distances. If $50\%$ is taken as the acceptable error threshold, the relative distance ($l/R$) ranges for the PFA and PWS techniques to satisfy this threshold are approximately $[0.02, 0.2]$ and $[0.2, 1]$.

\subsection{Influence of multiscale surface fluctuations on the disperse force}
\label{sec:multi}
As described in Section $\mathsection$\ref{sec:geoc}, the complex morphology of a real particle can be regarded as a superposition of surface fluctuations at different scales. The SEMD algorithm, which is an adaptive spatial filter, can be applied to sequentially filter surface fluctuations at different scales from fine to coarse.
With the average of the Gaussian curvatures of the surface maximums ($\overline{\kappa}_{c}$) taken as the scale characterization parameter, the influence of multiscale surface fluctuations on the disperse force can be quantitatively studied by calculating the force of each residual shape obtained by the SEMD algorithm.

\begin{figure}
\centering
\includegraphics[width=1.0\textwidth]{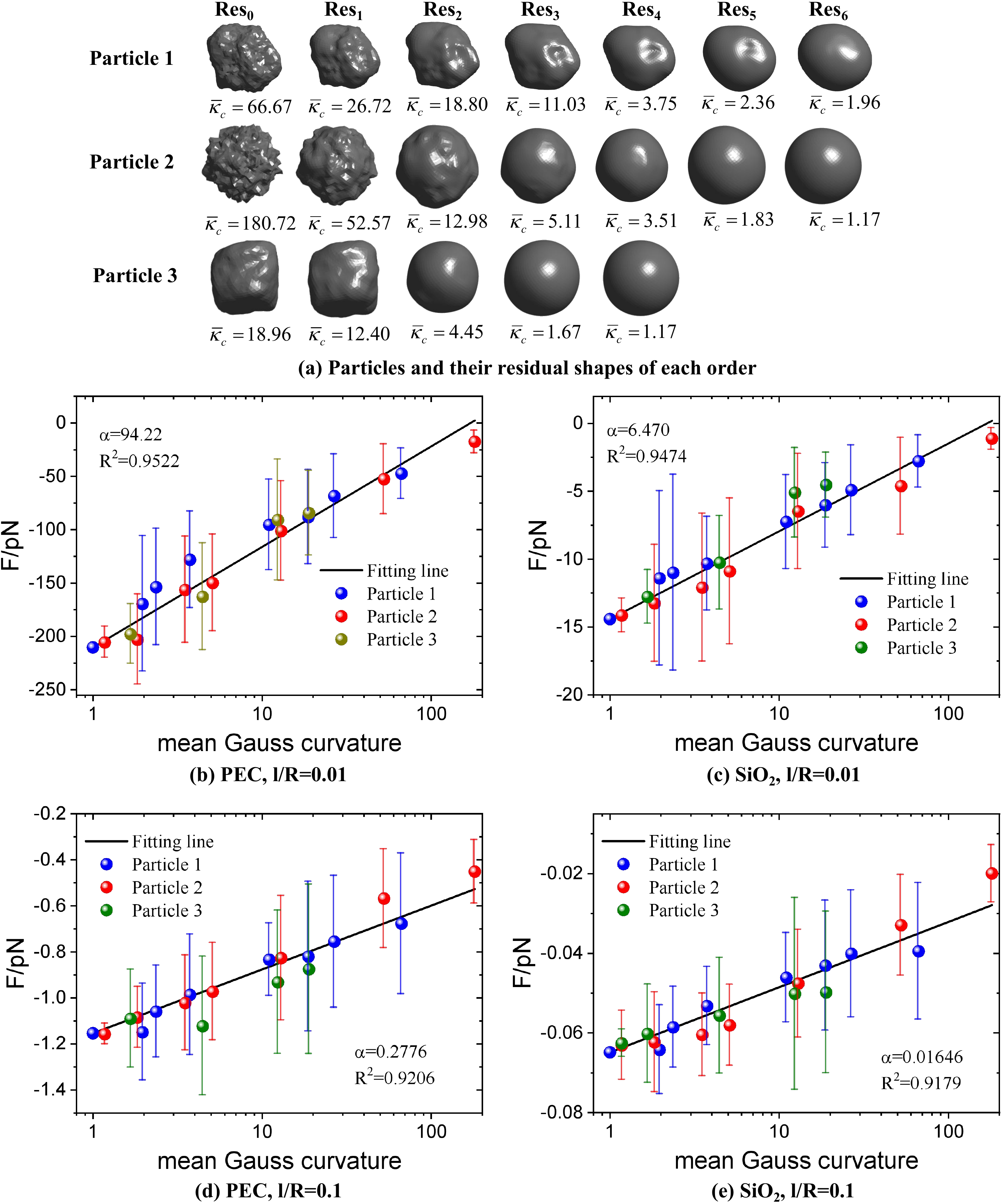}
\caption{Influence of multiscale surface fluctuations on the disperse force. (a) shows the residual shapes at each order of the decomposition of three selected particles. (b)-(e) show the variations of dispersion force with the average Gaussian curvature for the particles with surface fluctuations of different scale at different distances and materials. The colored dots represent the average of the cases generated by random rotation as shown in Fig. \ref{fig:3geoconf}, and the error bars refer to their standard deviations.}
\label{fig:multisurf}
\end{figure}

As shown in Fig. \ref{fig:multisurf}(a), three particles with large morphological differences were randomly selected, and their residual shapes at each order were obtained with the SEMD algorithm. The materials were set to PEC and \ch{SiO_2}, and the temperature was set to room temperature ($298.15 K$). The distances between the particles were set to $l/R = 0.01, 0.1$ and $1$.
As described in Section $\mathsection$\ref{sec:conf}, for cases with different distances, shapes and materials, a sufficient number of configurations must be obtained through random rotations. By testing different cases, we found that the mean and standard deviation of the dispersion force stabilized when the particles were randomly rotated more than 10 times; thus, we chose to randomly rotate the particles 20 times for each case.

Figs. \ref{fig:multisurf}(b)-(e) show the results for $l/R=0.01$ and $0.1$. When $l/R=1$, there is little difference in the dispersion forces of the different shapes, and these results are provided in the \textcolor{blue}{supplementary materials}.
The dispersion force decreases with increasing mean Gaussian curvature ($\overline{\kappa}_{c}$), and the mean value of the dispersion force and the logarithm of $\overline{\kappa}_{c}$ show a good linear correlation. Taking the dispersion force between spheres ($F_{ss}$) with equal volumes (i.e., the case of $\overline{\kappa}_{c}=1$) as the intercept, the linear fitting results in Figs. \ref{fig:multisurf}(b)-(e) all have $R^2$ values greater than 0.9. Therefore, the following relationship between the mean of the dispersion force and $\overline{\kappa}_{c}$ can be established:
\begin{eqnarray}
F(T,l/R)=\alpha \lg(\overline{\kappa}_{c})+F_{ss}(T,l/R),
\label{eq:fppkc}
\end{eqnarray}
where $\alpha$ is an empirical parameter that is determined by the distance and material.
The dispersion force gradually becomes insensitive to the morphology as the distance increases. When $l/R=0.01$, the variation range ($|F_{\max}|/|F_{\min}|$) is approximately $15$ times; when $l/R=0.1$, $|F_{\max}|/|F_{\min}|\sim 3$;
and when $l/R=1$, the variation range remains essentially constant.

Although the material significantly affects the magnitude of the dispersion force, it has little effect on the variation trend. Normalizing Eq. \eqref{eq:fppkc} by $F_{ss}(T, l/R)$ gives
\begin{eqnarray}
F_r=1+\beta \lg(\overline{\kappa}_{c}),
\label{eq:nfppkc}
\end{eqnarray}
where $F_r=F/F_{ss}$ and $\beta=\alpha/F_{ss}$. When $l/R=0.01$, $\beta=-0.4496$ for the PEC and $-0.4489$ for \ch{SiO_2}. When $l/R=0.1$, $\beta=-0.2406$ for the PEC and $-0.2537$ for \ch{SiO_2}. When $l/R=1$, $\beta\rightarrow 0$ for both materials. Thus, different materials have little effect on the value of $\beta$.

\begin{figure}
\centering
\includegraphics[width=0.5\textwidth]{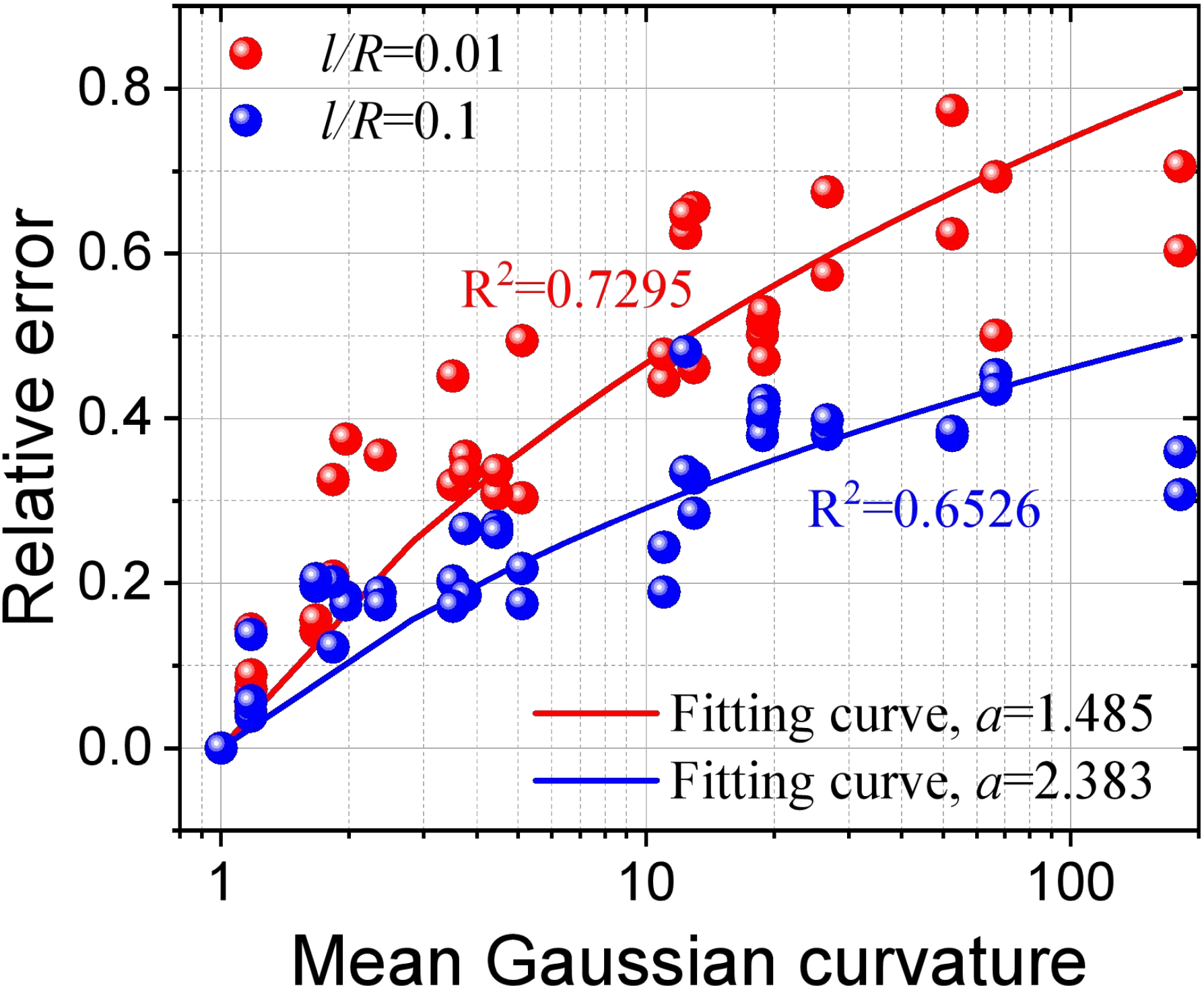}
\caption{Variation of relative error with mean Gaussian curvature at different distances}
\label{fig:multierr}
\end{figure}

The error bars in Fig. \ref{fig:multisurf} represent the standard deviations for each random rotation case. The ratio of the standard deviation to the mean value is taken as the relative error $\delta_r$, without distinguishing the material, and its variation with the morphology is shown in Fig. \ref{fig:multierr}. As the curvature increases, the relative error increases. Moreover, as the distance decreases, the relative error increases. Using $[\lg (1+\lg\overline{\kappa}_{c})]/a$ as the fitting function, the $R^2$ values exceed 0.65. The mean $\overline{\delta}_r$ and standard deviation $\sigma_{\delta r}$ were used to quantify the statistical characteristics of the relative error at different distances. When $l/R=0.01$, $\overline{\delta}_r\pm \sigma_{\delta r}=0.4296\pm 0.2055$. When $l/R=0.1$, $\overline{\delta}_r\pm \sigma_{\delta r}=0.2614\pm 0.1244$. With these values, we can determine to what extent the geometric anisotropy affects the discreteness of the dispersion force. In addition, the mean of the relative error $\overline{\delta}_r$ is positively correlated with the aforementioned $|\beta|$ defined in Eq. \ref{eq:nfppkc}.

By performing an exact numerical study, the influence law of multiscale surface fluctuations on the dispersion force was obtained quantitatively. In summary, the dispersion force decreases as the curvature increases, the shape sensitivity of the force decreases as the distance increases, and the material has little effect on the change trend. Taking the dispersion force between spheres with the same volume and material as a reference, the effect of the morphology and distance can be expressed as a simple empirical equation, as shown in Eq. \eqref{eq:nfppkc}.
As the irregularity in the shape increases and the distance decreases, the discreteness of the force increases due to geometric anisotropy, and the range of the variation of the relative error is shown in Fig. \ref{fig:multierr}.

\subsection{Multistage approximation of the dispersion force of real particles}
\label{sec:3levelc}
The nonadditivity algorithm is accurate but has a considerable computational cost.
When the distance is small, UV frequencies have substantial contributions; thus, many Matsubara frequencies need to be calculated for the force calculation to converge (e.g., when $R=1\mu m, l/R=0.01$, 500 orders are needed).
In practice, it is difficult to perform accurate calculations for different particles with various morphologies. Therefore, establishing a law and providing a prediction formula for the dispersion force based on the exact numerical study are both valuable.

\begin{figure}
\centering
\includegraphics[width=1\textwidth]{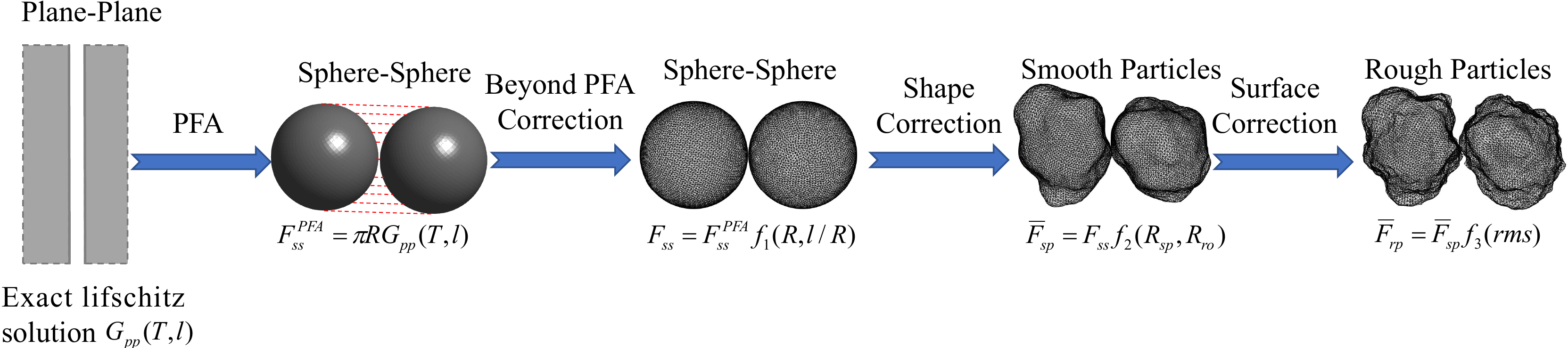}
\caption{Schematic diagram of the multistage correction of the real particle dispersion force, where $f_i()$ refers to the correction function of each stage to be determined.}
\label{fig:multicorr}
\end{figure}

Until now, an exact analytical solution of this force has been limited to the Lifshitz solution in the plane-plane configuration (as shown in Eq. \eqref{eq:pfa1}). According to the PFA method, the dispersion force in the sphere-sphere configuration can be expressed as a function of the radius and the energy in the plane-plane configuration:
\begin{eqnarray}
F_{ss}^{PFA}(T,l/R)=\pi R G_{pp}(T,l).
\label{eq:sspfa}
\end{eqnarray}
Here, we consider two spheres with the same radius $R$. Unfortunately, the PFA approach is effective only for limited distances and curvatures. As a result, the 'beyond PFA correction' was proposed and discussed \citep{Neto:2005roughness, Krause:2007experimental, Hartmann:2017plasma}. The basic idea of the beyond PFA correction is to multiply the PFA solutions by a correction factor according to the geometric descriptors.
Based on this idea, we propose a multistage correction method for predicting the dispersion forces of real particles, as illustrated in Fig. \ref{fig:multicorr}.
Eq. \eqref{eq:sspfa} gives the first PFA correction to the Lifshitz solution ($G_{pp}$) in the sphere-sphere configuration ($F_{ss}^{PFA}$), and the remaining correction stages are discussed below.

\emph{From $F_{ss}^{PFA}$ to $F_{ss}$:}
Previous studies \citep{Hartmann:2017plasma,Bimonte:2018beyond} have focused on this stage. According to the derivative expansion approach, the leading correction has the following form:
\begin{eqnarray}
\frac{F_{ss}}{F_{ss}^{PFA}}=1+\alpha_{ss}\frac{l}{R}+\cdots,
\label{eq:ssp2ss1}
\end{eqnarray}
where the coefficient $\alpha_{ss}$ is independent of $R$. Previous studies have indicated that the distance is mainly restricted to the range of $l/R\sim[0.1, 1]$ and that the dielectric function is restricted to the Drude and plasma models. As a result, the range of $\alpha_{ss}$ and the form of subleading corrections remain controversial.

\begin{figure}
\centering
\includegraphics[width=0.5\textwidth]{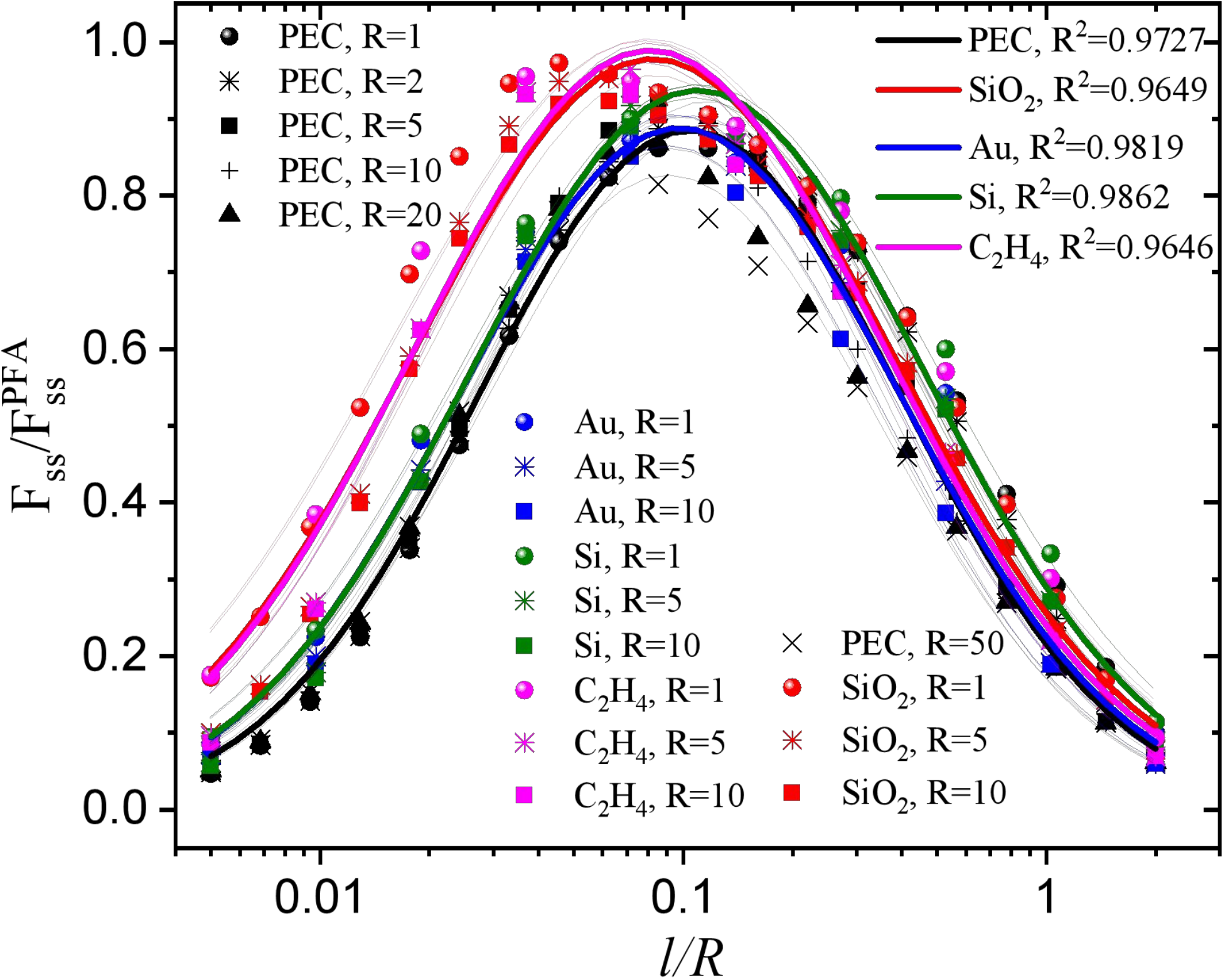}
\caption{Variation of the ratio of $F_{ss}$ (obtained by FSC) and $F_{ss}^{PFA}$ (obtained by PFA) with distance between two spheres with equal radius, and their corresponding curve fittings.}
\label{fig:sphcorr}
\end{figure}

Here, we considered a wider range of distances ($[0.005,2]$) with the FSC algorithm and more materials (PEC, \ch{Au}, \ch{SiO_2}, \ch{Si}, \ch{C_2H_4}) using the modified empirical harmonic oscillator model proposed by \citet{Moazzami:2021self}.
The calculation results are given in Fig. \ref{fig:sphcorr}. Taking the distances ($l/R$) on logarithmic coordinates, we found that the curves of $F_{ss}/F_{ss}^{PFA}$ show bell-shaped characteristics. Fitting with the Gaussian function, the $R^2$ can all reach above 0.97. Therefore, from $F_{ss}^{PFA}$ to $F_{ss}$, we use the Gaussian function in logarithmic coordinates as the new correction function:
\begin{eqnarray}
\frac{F_{ss}}{F_{ss}^{PFA}}=\alpha_{ss1}\exp\left[-\frac{(lx-\alpha_{ss2})^2}{2\alpha_{ss3}^2}\right],lx=\lg(l/R),
\label{eq:ssp2ss2}
\end{eqnarray}
in which, $\alpha_{ss1}$, $\alpha_{ss2}$ and $\alpha_{ss3}$ are parameters to be determined.
It can be seen that these parameters depend mainly on the type of material, and they are also related to $R$, but have little influence.
Therefore, the effect of $R$ is ignored in this study, and the values of these parameter for the five materials are provided in Table \ref{tab:ass} by fitting all radii together.

\begin{table}[!htpb]
\centering
\begin{ruledtabular}
\begin{tabular}{cccccc}
Materials & PEC  &  \ch{SiO_2} & \ch{Au} & \ch{Si} & \ch{C_2H_4} \\
\hline
$\alpha_{ss1}$ & 0.8851 & 0.9782 & 0.8885 & 0.9374 & 0.9895 \\
$\alpha_{ss2}$ & -0.9826 & -1.0844 & -1.0106 & -0.9607 & -1.0920 \\
$\alpha_{ss3}$ & 0.8623 & 0.9359 & 0.8615 & 0.8865 & 0.9171 \\
\end{tabular}
\end{ruledtabular}
\caption{Fitting parameters for different materials}
\label{tab:ass}
\end{table}

\emph{From $F_{ss}$ to $F_{sp}$:} This correction is the main work of this study. Eq.\eqref{eq:nfppkc} obtained in the Sec.$\mathsection$\ref{sec:multi} is an acceptable correction.
However, in practice, three-level descriptors are more commonly used to characterize the morphology, where sphericity and roundness represent the global form and local corner features, respectively.
As mentioned before, the curvature that can be accurately calculated at finite resolution is limited considering the computational load.
The SEMD-based three-level classification can filter out the influence of surface fluctuation to obtain 'smooth particles' with the same sphericity and roundness as the original morphology, and their curvature is in the range that can be accurately calculated. By calculating a large number of 'smooth particles' with different sphericity and roundness, the shape correction from $F_{ss}$ to $F_{sp}$ can be obtained.

\begin{figure}
\centering
\includegraphics[width=1\textwidth]{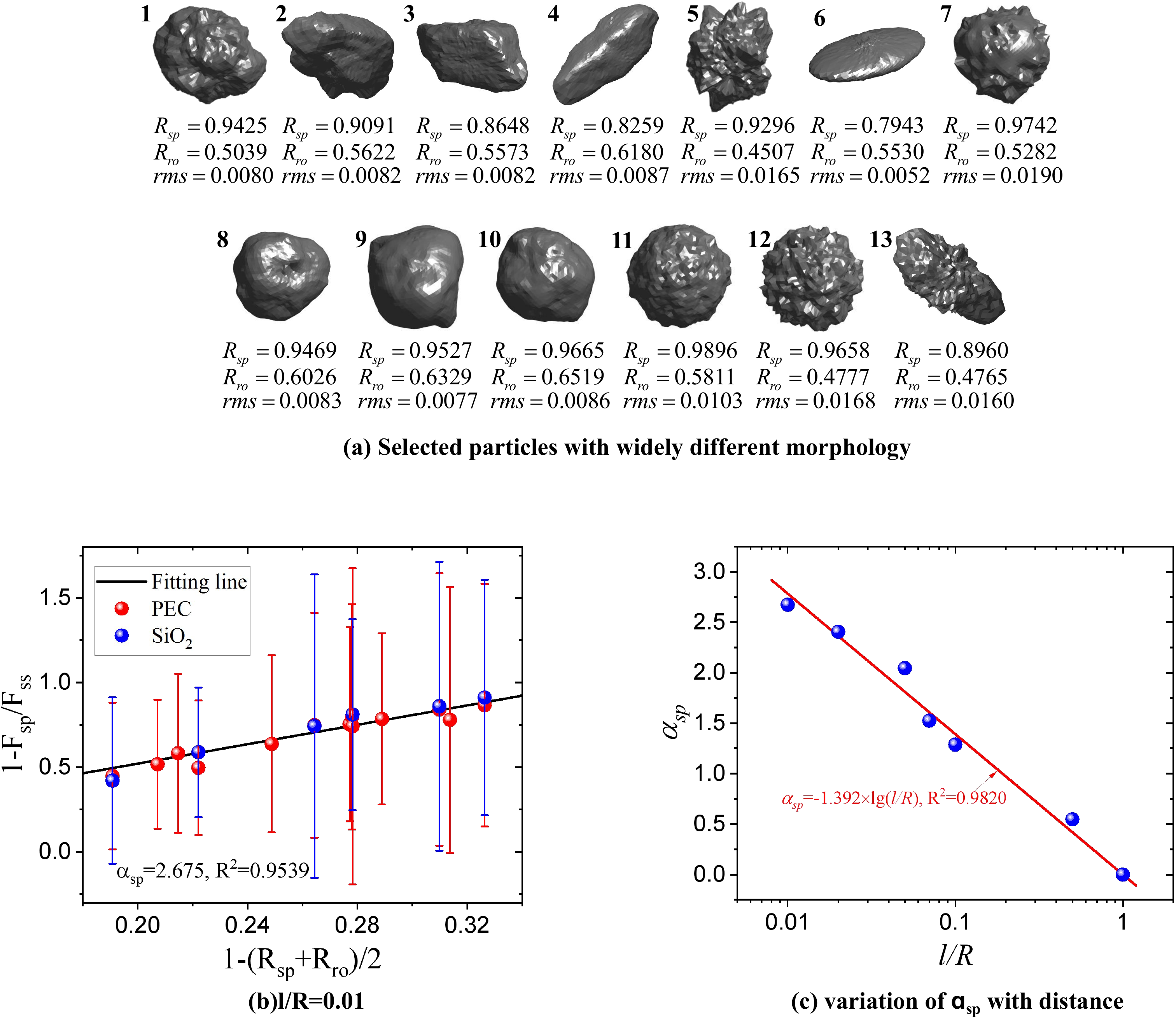}
\caption{Shape correction of dispersion force between arbitrarily shaped particles. (a) shows the widely different morphologies of the selected particles. (b) shows the excellent linearity between $1-F_{sp}/F_{ss}$ and $1-(R_{sp}+R_{ro})/2$, and the error bar refers to the relative error (standard deviation/mean) generated by the random rotation. (c) shows the relationship between paramater $\alpha_{sp}$ and relative distance $l/R$.}
\label{fig:shapecorr}
\end{figure}

Here, 13 particles with widely different morphology were chosen (as shown in Fig.\ref{fig:shapecorr}(a)), and the temperature was set to room temperature ($298.15K$). Same as the previous section, 20 random rotations are performed for each case. Taking the materials as PEC and \ch{SiO_2}, the results when $l/R=0.01$ are shown in Fig.\ref{fig:shapecorr}(b). It can be seen that the following relationship holds:
\begin{eqnarray}
\frac{F_{sp}}{F_{ss}}=1+\alpha_{sp}(\frac{R_{sp}+R_{ro}}{2}-1),
\label{eq:ss2sp}
\end{eqnarray}
where $\alpha_{sp}$ is the only one coefficient to be determined. The coefficient is not sensitive to the materials and depends mainly on the distance.
By calculating more different distances for the smooth particles of the PEC material, the relationship between the coefficient $\alpha_{sp}$ and the relative distance was obtained as shown in Fig.\ref{fig:shapecorr}(c).
As the distance increases, $\alpha_{sp}$ decays exponentially, i.e., the effect of shape on the force decays exponentially.
When $l/R>1$, it can be assumed that the shape has no effect on the force, and $F_{ss}$ can be used instead of $\overline{F}_{sp}$.

\emph{From $F_{sp}$ to $F_{rp}$:}
The correction at this stage is consistent with the roughness correction in previous studies \citep{Eichenlaub:2004roughness, Svetovoy:2015influence, Lamarche:2017method}.
The difference is that the influence of the surface fluctuations with smaller curvatures have been taken into account in the previous stages of this study, so that the correction at this stage can be controlled at the level of 'perturbation'.
It should be noted here that the descriptor of surface texture depends on the resolution of the morphological data.
The SEMD in this study decomposes the cross-linkage between surface and shape features of the particles, i.e., the resolution of the morphological data does not affect the shape descriptors, but does affect the surface descriptor.
As discussed before, the resolution calculated in this study is limited to 5000 surface triangular ($\overline{\kappa}_c<200$), and the roughness obtained by SEMD filtering is mainly concentrated between 0.007 and 0.009. Taking the average for all the cases, we get $\left\langle\left|F_{rp}-F_{sp}\right|/\left|F_{rp}\right|\right\rangle$=0.9794, which means that the error caused by surface texture is within $2.1\%$. Considering the limited range of roughness variation and the small error, the classical roughness perturbation correction formula \citep{Bordag:1995corrections} is directly adopted here, which can be expressed as
\begin{eqnarray}
\frac{F_{rp}}{F_{sp}}=1+\alpha_{rp}rms^2,
\label{eq:sp2rp}
\end{eqnarray}
where $\alpha_{rp}$ is the only one coefficient to be determined, which satisfies $\alpha_{rp}\propto (l/R)^{-2}$.

In summary, the dispersion force between complex shaped particles can be predicted from the exact Lifshitz solution by the multi-stage corrections.
By combining Eq.\eqref{eq:sspfa}, \eqref{eq:ssp2ss2}, \eqref{eq:ss2sp} and \eqref{eq:sp2rp}, The final form of the prediction formula can be expressed as follows:
\begin{eqnarray}
\begin{aligned}
F_{rp}={\color{red}\pi R} {\color{blue}G_{pp}(T,l)} {\color{olive}\alpha_{ss1}\exp\left[-\frac{(\lg(l/R)-\alpha_{ss2})^2}{2\alpha_{ss3}^2}\right]}\\
&\times{\color{magenta}\left[1+\alpha_{sp}\left(\frac{R_{sp}+R_{ro}}{2}-1\right)\right]}{\color{cyan}(1+\alpha_{rp}rms^2)},
\end{aligned}
\label{eq:com}
\end{eqnarray}
in which, the blue part refers to exact Lifshitz solution, the red part refers to PFA correction, the olive part refers to beyond PFA correction of sphere-sphere configuration, the magenta part refers to shape correction and the cyan part refers to surface correction. Factors influencing the dispersion force include geometry, material and temperature. Among them, the influence of geometry were taken into account in the multistage corrections, while the influence of material and temperature were reflected in the exact Lifshitz solution.

To verify the validity of formula \eqref{eq:com}, more examples of different materials with different morphologies at different temperatures are added, and
the calculated values by FSC and predicted values by formula \eqref{eq:com} of all cases are shown in Table \ref{tab:com} for comparison.
It can be found that the prediction error of the mean dispersion force is basically within $15\%$ for the particles with complex morphology. Compared with the sphere-sphere PFA ($\pi R G_{pp}$) commonly used in engineering, the error is reduced significantly, but the computational complexity is not increased. This indicates that the formula \eqref{eq:com} is a reasonable prediction. one can conveniently predict the mean dispersion force with the knowledge of materials, temperature and geometry descriptors, and the corresponding program can be found at \url{https://github.com/yfliu088/vdw-casimir}.

\begin{table}[!htpb]
\centering
\begin{ruledtabular}
\resizebox{\linewidth}{!}{
\begin{tabular}{ccccccccccc}
ID & Materials  &  Temp. & Radius & Distance & Sphericity & Roundness  &  $\pi R G_{PP}$ & Prediction & FSC & Error \\
 &    &  ($K$) & ($\mu m$) & ($l/R$) &  &   &  ($pN$) & ($pN$) & ($pN$) & Relative \\
\hline
1  & PEC	  & 170.7 &	1.191  &	0.0110 &	0.9199 &	0.5149 	& -701.5	& -40.19    & -35.18	&     0.1424 \\
2  & \ch{SiO_2}  & 217.2 &	3.031  &	0.0944 &	0.9153 &	0.4807 	& -0.0112   & -6.19e-3	& -6.16e-3	& 0.0056     \\
3  & \ch{Au}	  & 83.84 &	3.159  &	0.0162 &	0.8976 &	0.5116 	& -12.87	& -1.336    & -1.227	&     0.0894 \\
4  & \ch{Si}	  & 209.7 &	5.317  &	0.1313 &	0.9241 &	0.5987  & -6.49e-3	& -4.27e-3	& -4.87e-3	& 0.1250     \\
5  & \ch{C_2H_4}  & 51.16 &  9.673 & 	0.0987 &	0.9119 &	0.6299 	& -7.69e-4	& -5.12e-4	& -5.74e-3	& 0.1092     \\
6  & PEC	  & 84.25 &	9.735  &	0.2078 &	0.8854 &	0.5861 	& -1.50e-3	& -8.82e-4	& -1.05e-3	& 0.1609     \\
7  & \ch{SiO_2}  & 131.1 &	9.086  &	0.0741 &	0.7944 &	0.5952 	& -3.16e-3	& -1.60e-3	& -1.82e-6	& 0.1188     \\
8  & \ch{Au}	  & 109.0 &  4.639 & 	0.0240 &	0.8076 &	0.7063 	& -2.538	& -0.6207   & -0.5373	&     0.1550 \\
9  & \ch{Si}	  & 158.3 &	7.997  &	0.1054 &	0.9243 &  0.5281 	& -2.589	& -0.2364   & -0.2445 	& 0.0331     \\
10 & \ch{C_2H_4}  & 8.192 &	2.101  &	0.0371 &	0.8937 &	0.5403 	& -0.2872 	& -0.1084	& -0.1004  	& 0.0797     \\
\end{tabular}}
\end{ruledtabular}
\caption{Comparison between the prediction by \eqref{eq:com} and calculation by FSC. Considering the small impact, the roughness were uniformly taken as 0. The temperatures were chosen randomly between $[0,300] K$, the Radii were chosen randomly between $[1,10] \mu m$, and the relative distances were chosen randomly between [0.01,0.5]. The particle morphology corresponding to each ID were provided in the \textcolor{blue}{supplementary materials}.}
\label{tab:com}
\end{table}

Considering the computational load, the resolution of the calculations in this study is limited, so the above conclusions are guaranteed to be valid for a certain curvature range.
However, due to the introduction of the SEMD method, we decouple and interconnect the shape and the surface.
That is, up to the stage of shape correction, our conclusions are universal, regardless of the resolution of the real morphological data.
Meantime, the last remaining surface correction is guaranteed at the perturbation level.
Therefore, for cases where the resolution exceeds the computational limits of this paper, the results of this study can be used for shape correction, and the existing perturbation approximation theory can be used for the last roughness correction.

\section{Conclusion}
\label{sec:con}
Geometric analyses and accurate calculations of the dispersion forces for particles with various complex shapes were provided, and the conclusions about the significance and reach of this work can be summarized as follows:

\begin{enumerate}
\item By comparing our approach with the optimal versions of the additivity algorithms, we found that the PFA and PWS approaches tend to overestimate the dispersion force for particles with complex shapes. With an error threshold of $50\%$, the relative distance ($l/R$) range for the PFA and PWS methods to satisfy this threshold are approximately $[0.02, 0.2]$ and $[0.2, 1]$.
\item We performed an exact numerical study for shapes with different curvatures obtained by an adaptive spatial filter, revealing that the average value of the dispersion force and the logarithm of the curvature show an excellent linear correlation.
\item Based on the results of large-scale nonadditivity calculations and multistage corrections, a convenient formula for predicting the dispersion force between complex-shaped particles from the exact Lifshitz solution was established.
\end{enumerate}

Our work extends the study of the dispersion forces between micro/nanoparticles with complex morphologies considering nonadditivity, improves the understanding of cohesion, and lays the foundation for a more reasonable contact model between irregular particles.

\section*{Acknowledgments}
This study was funded by the National Natural Science Funding of China (Nos. 52104141, 12172230, U2013603, and 51827901) and the Department of Science and Technology of Guangdong Province (No. 2019ZT08G315).

\section*{Conflicts of Interest}
The authors declare that there are no conflicts of interest regarding the publication of this article.

\appendix
\renewcommand\thefigure{\Alph{section}\arabic{figure}}
\section{Discretization study}
\label{sec:appA}
\setcounter{figure}{0}
The discretization study is applied to particles of different curvatures. Residual shapes of different SEMD order for the same particle are chosen. The distance is set to $l/R=0.1$, the material is set to PEC, the temperature is set to room temperature ($298.15K$), and the calculation results for different grid resolutions $Dt$ are shown in Figure 2. The relative error $e_r$ in the figure is calculated by the following equation
\begin{eqnarray}
e_r=\frac{\left|F_{Dt}-F_{\infty}\right|}{\left|F_{\infty}\right|},
\label{eq:er}
\end{eqnarray}
where $F_{Dt}$ refers to the force calculated when the resolution is taken as $Dt$, and $Dt$ here refers to the number of surface triangular element. The $F_{\infty}$ refers to the exact value of the force when the resolution approaches infinity, which can obtained by the Richardson extrapolation:
\begin{eqnarray}
F_\infty=\frac{(Dt_2/Dt_1)^2F_{Dt_2}-F_{Dt_1}}{(Dt_2/Dt_1)^2-1},
\label{eq:inf}
\end{eqnarray}
where $Dt_2$ and $Dt_1$ denote two different resolutions.
\begin{figure}
\centering
\includegraphics[width=0.5\textwidth]{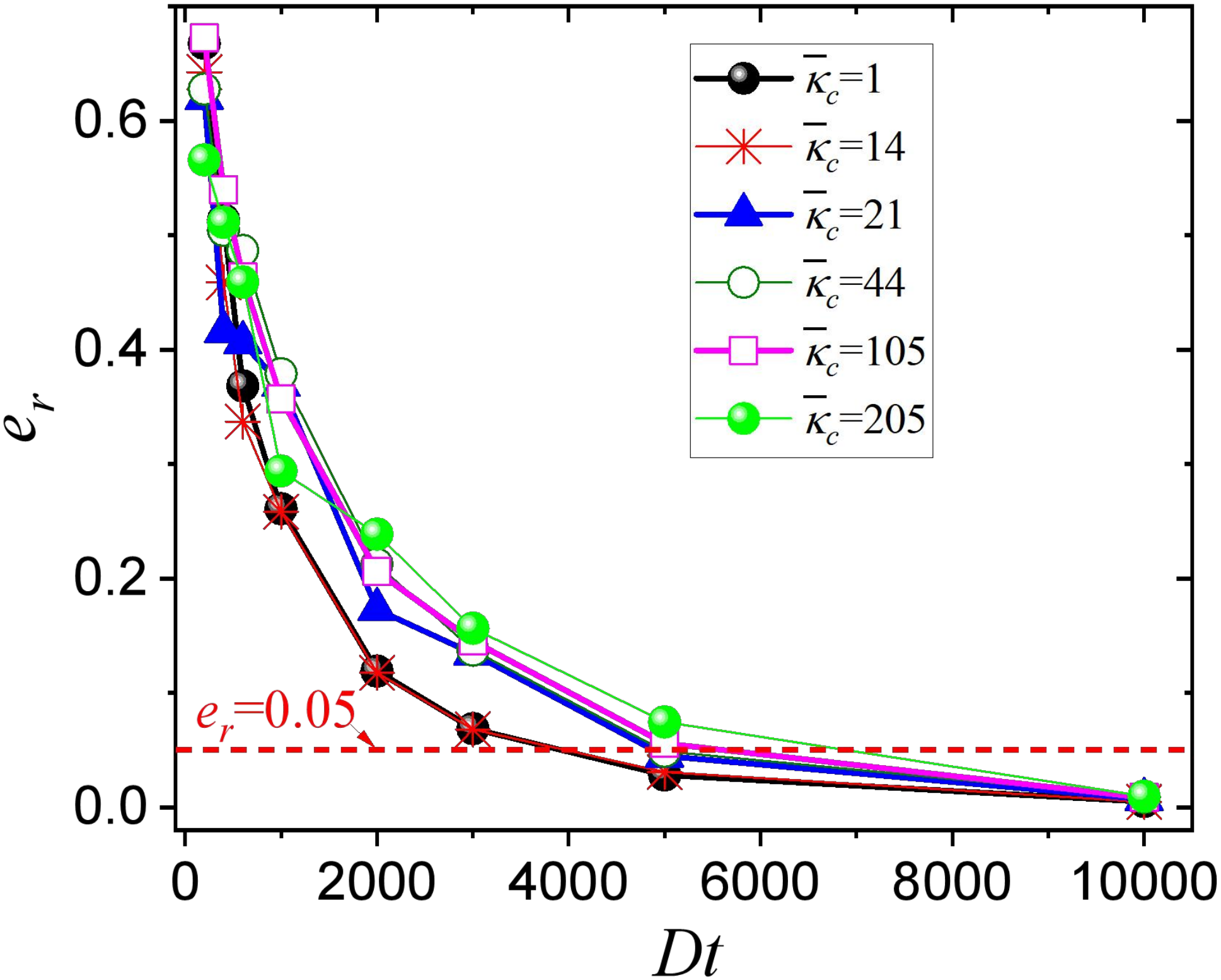}
\caption{Variation of the relative error with resolution.}
\label{fig:res}
\end{figure}
Considering the computational efficiency, the resolution is set to 5000 surface triangular cells. Using 0.05 as the acceptable threshold of relative error, it can be seen that the accuracy can be guaranteed when $\overline{\kappa}_c<100$.

\bibliographystyle{apsrev}
\bibliography{draft_ev_v7}

\end{document}